\documentclass[a4paper]{jpconf}
\usepackage{graphicx}
\begin{document}
\title{The ALICE electromagnetic calorimeter high level triggers}

\author{F. Ronchetti$^1$, F. Blanco$^2$, M. Figueredo$^3$, A.G. Knospe$^4$ and L.~Xaplanteris$^4$ for the ALICE HLT Collaboration}

\address{$^1$INFN, Laboratori Nazionali di Frascati, CP 13, 00044 Frascati (Roma), Italy}
\address{$^2$The University of Houston, 4800 Calhoun Road, Houston, TX 77004-2693, USA}
\address{$^3$Universidade de S\~{a}o Paulo, Instituto de F\'{i}sica, Rua do Mat\~{a}o, 187, S\~{a}o Paulo, 05508-090, Brazil}
\address{$^4$The University of Texas at Austin, 1 University Station, C1600 Austin, TX 78712-0264, USA}

\ead{$^1$federico.ronchetti@lnf.infn.it}

\begin{abstract}
The ALICE (A Large Ion Collider Experiment) detector yields a huge sample of data from different sub-detectors. 
On-line data processing is applied to select and reduce the  volume of the stored data. 
ALICE applies a multi-level hardware trigger scheme where fast detectors are used to feed a three-level (L0, L1, and L2) deep chain.
The High-Level Trigger (HLT) is a fourth filtering stage sitting logically between the L2 trigger and the 
data acquisition event building. 
The EMCal detector comprises a large area electromagnetic calorimeter that extends the momentum measurement of photons and neutral mesons up to $p_T=250$~GeV/c, 
which improves the ALICE capability to perform jet reconstruction with measurement of the neutral energy component of jets.
An online reconstruction and trigger chain has been developed within the HLT framework to sharpen the EMCal hardware triggers, 
by combining the central barrel tracking information with the shower reconstruction (clusters) in the calorimeter. 
In the present report the status and the functionality of the software components developed for the EMCal HLT 
online reconstruction and trigger chain will be discussed, as well as  preliminary results from their commissioning
performed during the 2011 LHC running period.
\end{abstract}

\section{Introduction}
The ALICE experimental setup is read out at an overall data rate of up to 25 GByte/s. Thus
on-line data processing must be applied in order to reduce the data volume.
ALICE applies a multi-level hardware scheme where fast detectors are used to feed the Level-0/Level-1 (L0/L1) hardware trigger sequence,
followed by a Level-2 (L2)  trigger accept decision in the Central Trigger Processor available to provide past/future protection against pileup events. 
At the end of this chain, a more refined filtering stage is introduced: the High-Level Trigger (HLT), 
which is able to reduce the volume of the data stream to permanent storage by one order of magnitude.
The HLT layer is designed to perform complex event selection functions
via fast reconstruction algorithms in order to provide trigger
decisions, select Regions-of-Interest, and compress the data.
In addition to event selection and triggering tasks, the HLT produces online detector performance 
monitoring information and is able to perform on-line calibrations. 
The HLT on-line software components run within the publisher-subscriber 
data transport framework. In this context, EMCal specific processing components have been developed to perform 
reconstruction, monitoring, and triggering.

\section{Overview of the EMCal hardware triggers}
After a successful L0 and consecutive L1 and L2  accept trigger sequence, 
the HLT farm receives a copy of the raw data from each subsystem front-end and trigger electronics.
Specifically, the EMCal\cite{emc} Front End Electronics (FEE) provides to the HLT chain a full copy 
of the raw data. In addition, hardware trigger primitives 
(such as the list of the calorimeter cell clusters satisfying the trigger condition) produced by
the EMCal  Summary Trigger Unit  processor  (STU) are recorded along with the raw data and
are also provided to the HLT.

In {\it pp} collisions, the EMCal provides trigger input at L0 using a low threshold (2 GeV) 
to trigger on events with EMCal activity (electrons and photons) without bias from other trigger detectors.
The L0 trigger for showers is provided by the fast analog $2\times 2$ 
tower sums, also know as FastORs, and passed to a nearby Trigger Region Unit (TRU) where they are digitized.
The digitized $2\times2$ signals are then summed in the FPGA of the TRU in overlapping ($4\times 4$) tower regions 
and then compared to a programable threshold to generate the L0 single shower trigger. 

A single cluster trigger can be effectively generated by the leading particle of a jet 
(usually from a hard $\pi^0$). On the other hand, this trigger is significantly biased as a jet trigger since the leading particle does not carry the full jet energy. 
In addition, the bias is exacerbated in heavy ion collisions, when the leading particle is emitted deep inside the fireball and may
lose energy and correlation with the other particles of the jet by traversing a large fraction of the QCD medium
(an effect known as jet quenching\cite{vi}).
Since a single EMCal trigger unit (TRU) sees only a  fraction of the entire 
calorimeter acceptance, it is not suitable to generate a full jet trigger.
To overcome this limitation,  the digitized FastORs signals from all TRUs are passed to the STU FPGA 
which performs the integration over a programmable size window (denoted as jet patch) scanning over the entire EMCal
acceptance.
This L1 jet trigger provides a less biased trigger on high $E_t$ jets~\cite{a1,a2}
by integrating the electromagnetic energy over a large phase space area to trigger on a significant fraction of the total jet energy, 
 
The  jet trigger decision is driven by a multiplicity-dependent threshold.
For heavy ion collisions, a fixed jet energy threshold (with a value set to discriminate jets from the underlying energy in a jet patch region tuned for central events)
would have an unacceptably high threshold for jet signals from peripheral events. 
The solution implemented provides the collision multiplicity information signal
from a forward ALICE detector (V0) to the EMCal jet trigger unit (STU), to enable a centrality-dependent trigger threshold 
(recomputed on an event-by-event basis) to maintain an approximately uniform jet trigger efficiency across event centralities.

\section{The EMCal HLT online chain}

The EMCal L0 or L1 hardware trigger decisions provide the input 
for a dedicated on-line event processing chain running on the HLT cluster,
where further refinement based on criteria using the full event reconstruction 
information is performed.
In fact, the detector optical link transports the raw data to the 
Read-Out Receiver Card (RORC) in the local data collector of the data acquisition system, 
which sends a complete copy of the readout 
to a set of specialized nodes in the HLT cluster (FEP or Front End Processors).
Each FEP node is equipped with RORC cards in analogy to
the collector nodes used by the data acquisition.
The FEP nodes are physically linked to the detector hardware and reflect the geometrical
partitioning of each ALICE sub-system. 
Specifically, the EMCAL is composed of 10 full-size super-modules of 1152 channels 
and 2 reduced-size super-modules of 384 channels, for a total of 12 288 channels,
covering an azimuth $\Delta\phi=110\deg$ and a pseudo-rapidity $-0.7\le\Delta\eta\le0.7$.
The 10 full-size super-modules are read out using 2 Read-Out Control Units (RCUs) for a total
of 20 optical links running into the HLT FEPs. 
The reduced-size super-modules were installed prior to the 2012 LHC run and are not discussed in the present report.
In addition to the 20 links from the super-module readout, the HLT receives also a copy of the L0/L1 trigger data stream 
via an additional optical link from the EMCal jet trigger unit (STU) data collector.
The different stages of data processing are then performed by the 
software analysis chain executed on the HLT cluster:
a set of general purpose nodes (Computing Nodes or CNs) 
perform the higher level operations on the data streams which have been
already pre-processed on the FEPs at the lower level.
The EMCal software components form a specialized sub-chain 
executed at run time together with all other ALICE sub-systems
participating in the HLT event reconstruction. 

\begin{figure}[ht]
\includegraphics[width=23pc]{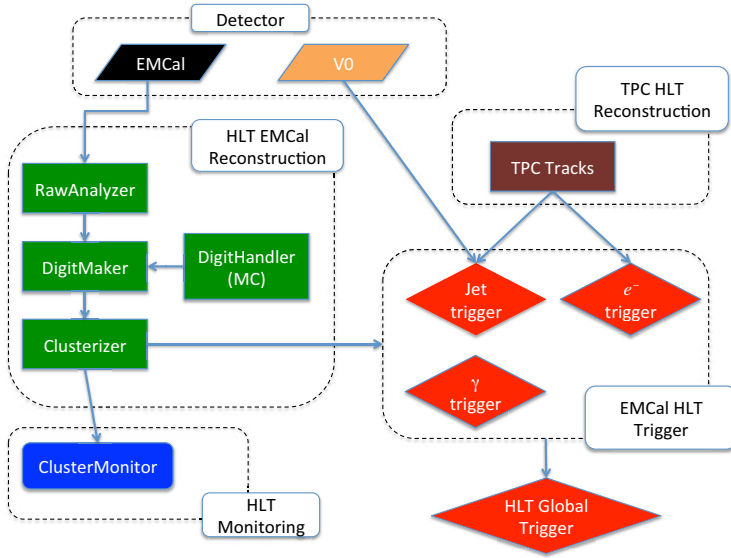}
\begin{minipage}[b]{14pc}\caption{\label{f1} Functional diagram of the EMCal online reconstruction components (signal processing, data structure makers, and clusterizers) shown in green. 
The EMCal chain is fed by the detector raw data. Trigger components are shown in red. EMCal-specific triggers operate on the calorimeter clusters
and perform TPC track-matching when needed (electron and jet triggers). Monitoring components are shown in blue and live in a separate monitoring chain.
The EMCal triggers are evaluated within the Global Trigger which is aware of the full HLT trigger logic of the other ALICE detectors.  
}
\end{minipage}
\end{figure}

The functional units of the EMCal HLT online chain are presented in Figure \ref{f1} where
the online reconstruction, monitoring, and trigger components 
are shown together with their relevant data paths.
The lower-level EMCal online component ({\it RawAnalyzer}) is fed by the detector front end electronics
and performs signal amplitude and timing information extraction.
Intermediate components ({\it DigitMaker}) use this information to 
build the digitized data structures needed for the clusterizer components to operate on the cell signals. 
Alternatively, the digitized signals can be generated via monte carlo simulations ({\it DigitHandler}).

At the top of the EMCal reconstruction chain, the digits are summed by the {\it Clusterizer} component to produce the cluster data structures. 
The calorimeter clusters are then used to generate the different kinds of EMCal HLT trigger information:
a single shower trigger ($\gamma$) with no track matching, an electron trigger using the matching with a corresponding TPC track,
and a jet trigger also using the TPC tracks information and the V0 multiplicity dependent threshold.

The trigger logic generated by the EMCal chain is evaluated 
(together with the outputs of the HLT trigger components coming from other ALICE detectors)
within the HLT Global Trigger which produces the final high level decision based on the reconstructed event.
The ALICE data acquisition system will then discard, accept or tag the event according to the HLT decision.

For performance and stability reasons, the full on-line HLT chain contains only analysis 
and trigger components. On the other hand, monitoring components typically make 
heavy use of histogramming packages and ESD objects, hence they are kept in a separate chain.
The isolation of the monitoring from the reconstruction chain gives additional robustness since 
a crash in a monitoring component will not affect the reconstruction chain and the data taking.

\section{Reconstruction components}

As shown in Figure \ref{f1} the EMCal HLT analysis chain provides all the 
necessary components to allow the formation of a trigger
decision based on full event reconstruction. The following 
subsections are devoted to a detailed discussion of each processing 
stage, starting from the most basic, i.e. signal extraction, to 
the highest stage: the HLT trigger decision.

\subsection{RawAnalyzer}
The {\it RawAnalyzer} component extracts energy and timing information for each calorimeter cell. 
Extraction methods implemented in the offline code (AliRoot) typically 
use least squares fitting algorithms, and cannot be used in online processing for 
performance reasons. 
Conversely, the HLT signal extraction is done without need of fitting 
using two possible extraction methods. The first method, referred to as {\it kCrude},
simply produces an amplitude using the difference between the maximum and the minimum values
of the digitized time samples and associates the time bin of the maximum as the signal arrival time.
The {\it kCrude} method was used during the 2011 data taking: it has the advantage
of being extremely fast and fully robust since no complex algorithms are used. On the other hand,
it produces a less accurate result than the processing of the full signal shape.
An alternative method ({\it kPeakFinder}\cite{peak}) evaluates the amplitude 
and peak position as a weighted sum of the digitized samples. This approach is
not as fast as {\it kCrude} but is a few hundred times faster than least squares fitting.

\subsection{DigitMaker} 
The {\it DigitMaker} component essentially transforms the raw cell signal amplitudes
produced by the {\it RawAnalyzer} into digit structures by processing the cell coordinates 
and by the application of dead channel maps and the appropriate gain factors (low and high-gain).

\subsection{Clusterizer} 
The {\it Clusterizer} component merges individual signals (digits) of adjacent cells
into  structures called clusters.
At transverse momenta $p_T>1$~GeV/c most of the clusters are associated to electromagnetic 
showers in EMCal  from $\pi^0$ and $\eta$ mesons decays. 
Other sources of electromagnetic showers are direct photons 
and electrons from semi-leptonic decays of $c$ and $b$ hadrons.
Since the typical cluster size in the EMCal can vary according to the detector occupancy due to shower 
overlap effects, which are much different for {\it pp} and heavy-ion collisions, 
clustering algorithms with and without a cutoff on the shower size are available (both in offline and in the HLT)  
to optimize the cluster reconstruction for the different cases. 
Events originating from {\it pp} collisions tends to generate 
smaller, spherical and well-separated clusters in the EMCal, at least up to 10 GeV/c. 
At higher transverse momenta, overlapping of the showers requires a shape analysis 
to extract the single shower energy. 
Above 30 GeV/c the reconstruction can be performed only with more sophisticated 
algorithms such as isolation cuts to identify direct photons.

\begin{figure}[ht]
\begin{minipage}{18pc}
\includegraphics[width=18pc]{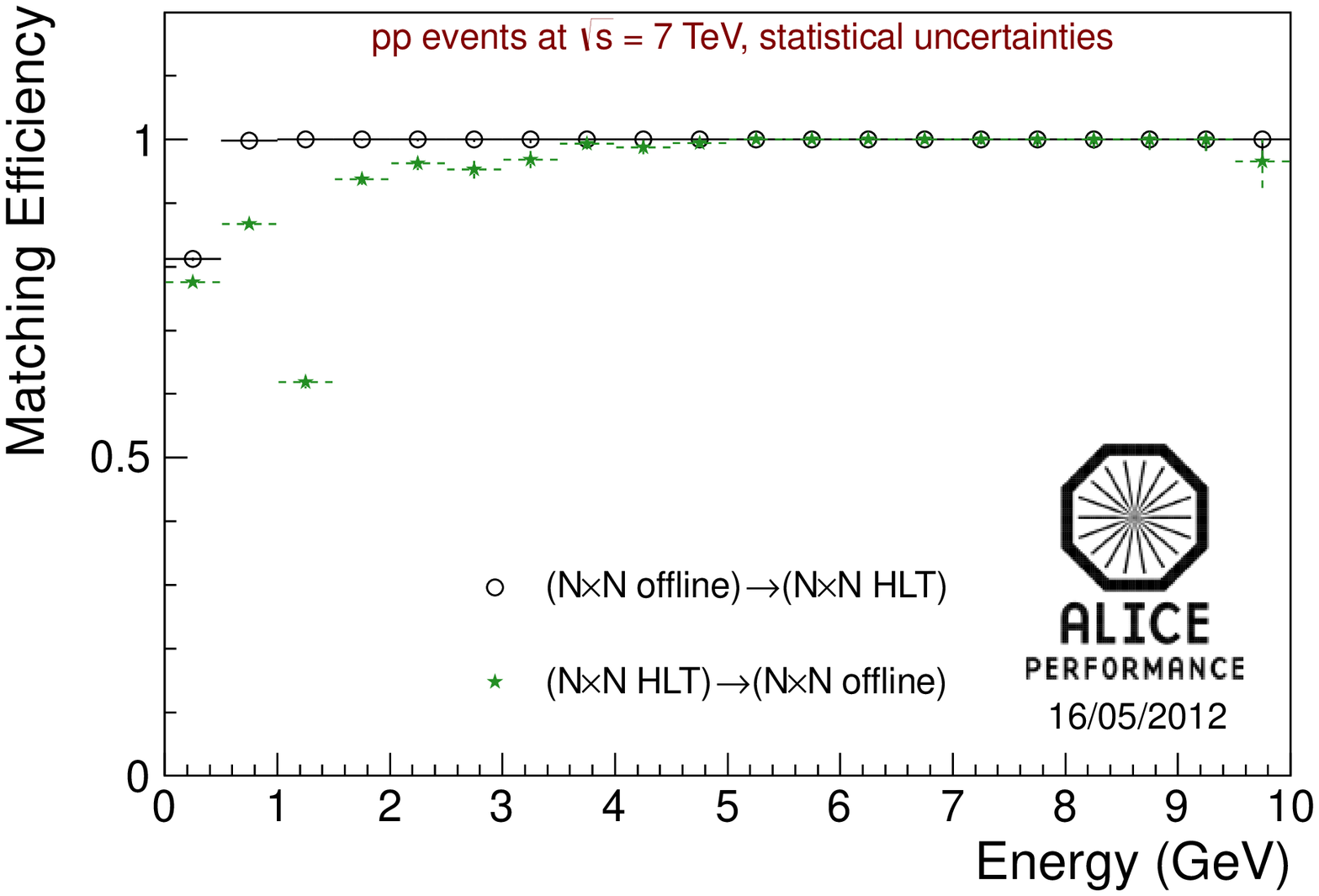}
\caption{\label{f5} Reconstruction efficiency for the $N\times N$ algorithm  (cutoff) in offline and HLT. The notation $(A) \rightarrow (B)$ indicates
the fraction of clusters found using method A that are also found using method B (data from run 154787, period LHC11c).}
\end{minipage}\hspace{2pc}
\begin{minipage}{18pc}
\includegraphics[width=18pc]{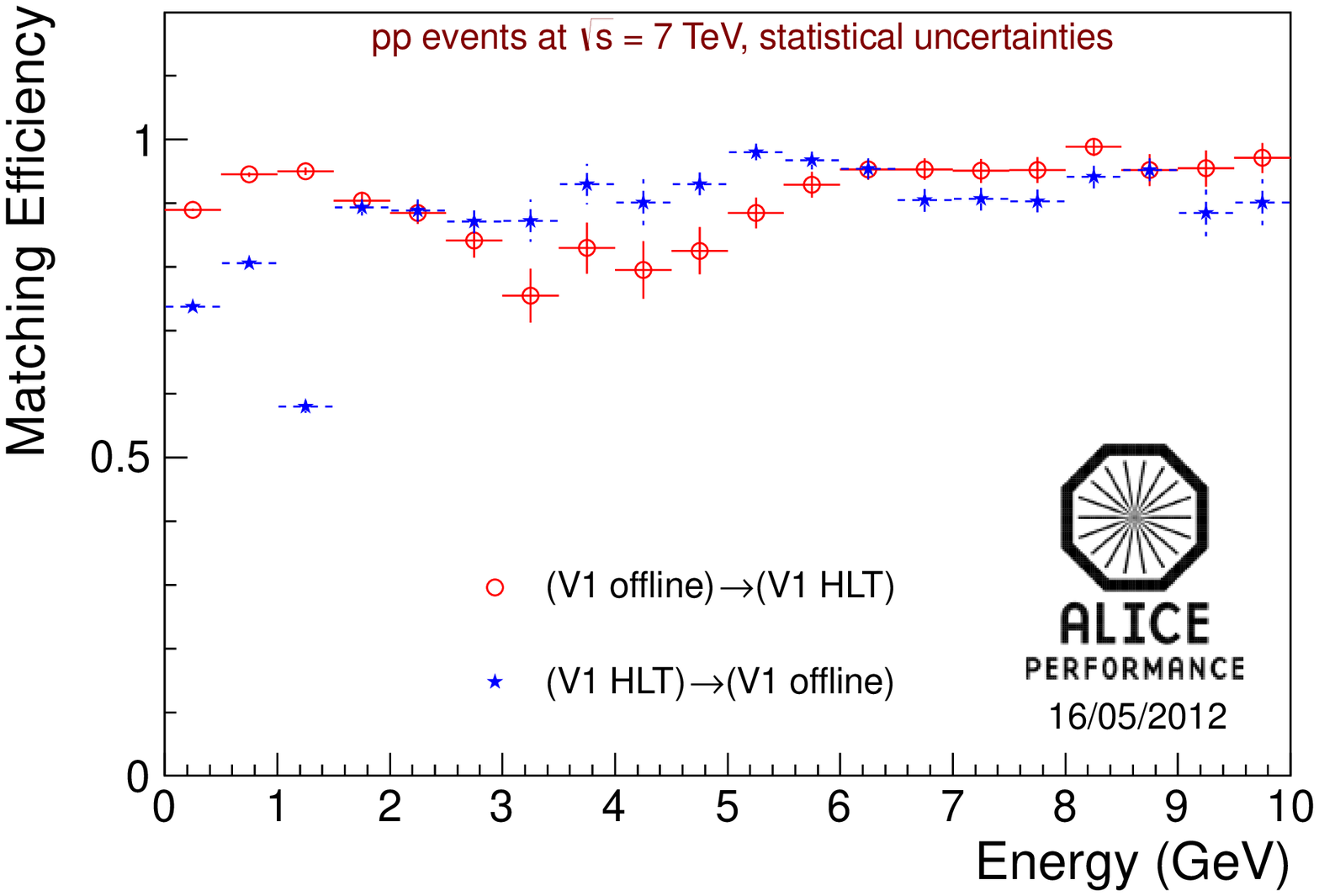}
\caption{\label{f6}Reconstruction efficiency for the $V1$ algorithms (no cutoff) in offline and HLT. The notation $(A) \rightarrow (B)$ indicates
the fraction of clusters found using method A that are also found using method B (data from run 154787, period LHC11c).}
\end{minipage} 
\end{figure}

The identification of an isolated single electromagnetic cluster in the EMCal can be performed using different
strategies: summing up all the neighboring cells around a seed-cell over threshold until no more cells are found 
or adding up cells around the seed until the number of clustered cells reaches the predefined cutoff value.

The first approach is more suitable for an accurate reconstruction. A further improvement to this 
clustering algorithm would be the ability to unfold overlapping clusters as generated from the 
photonic decay of high-energy neutral mesons, however this procedure usually requires computing intensive fitting algorithms. 

Such performance penalty must be avoided in the online reconstruction so the cutoff
technique is preferred.  In the EMCal HLT reconstruction a cutoff of 9 cells is used (according to the 
geometrical granularity of the single cell size), so the clusterization is performed into a 
square of $3\times3$ cells. The cutoff and non-cutoff algorithms are referred to as $N\times N$ and $V1$, respectively.

In {\it pp} collisions the response of the two methods is very similar since the majority of clusters are well separated, while  
in {\it PbPb} collisions, especially in central events, the high particle multiplicity requires the use of the cutoff (or unfolding in offline) 
to disentangle the cluster signals from the the underlying event to avoid the generation of  artificially large clusters.

The quality of the EMCal online clusterizer algorithms implemented in the HLT chain were checked against offline, 
as shown in Figures \ref{f5} and \ref{f6} where it can be seen that the performance is in a reasonable agreement in all cases.
The low point at 1.25 GeV is due to bad towers, which are assigned an energy of 1 GeV.  
Bad clusters are removed in later stages of the analysis, but that is not yet reflected in Figures \ref{f5} and \ref{f6}.  
This effect leads to an excess of clusters that are found by the HLT clusterizer, but not by the offline clusterizer.\\

Since the EMCal HLT reconstruction is mainly targeted for triggering, a small penalty in the accuracy of the energy reconstruction of the clusters is accepted 
as a trade off in favor of faster performance, and for this reason the cutoff clustering method was used, especially in {\it PbPb} collisions.

\section{Trigger components}

The online HLT chain is capable of producing trigger decisions based on full
event reconstruction. In terms of EMCal event rejection the following relevant trigger observables
have been implemented:

\begin{itemize}
\item neutral cluster trigger
\item electron and jet trigger
\end{itemize}

\subsection{Cluster trigger} 
The single shower triggering mode is primarily targeted to trigger on photons and neutral mesons.
In all collision systems, the high level trigger post-filtering can improve  
the hardware L0 and L1 trigger response by using the current bad channels map information
and calibration factors (which could be recomputed directly in the HLT).

\subsection{Electron trigger}
\begin{figure}[hb]
\begin{minipage}{18pc}
\includegraphics[width=17.5pc]{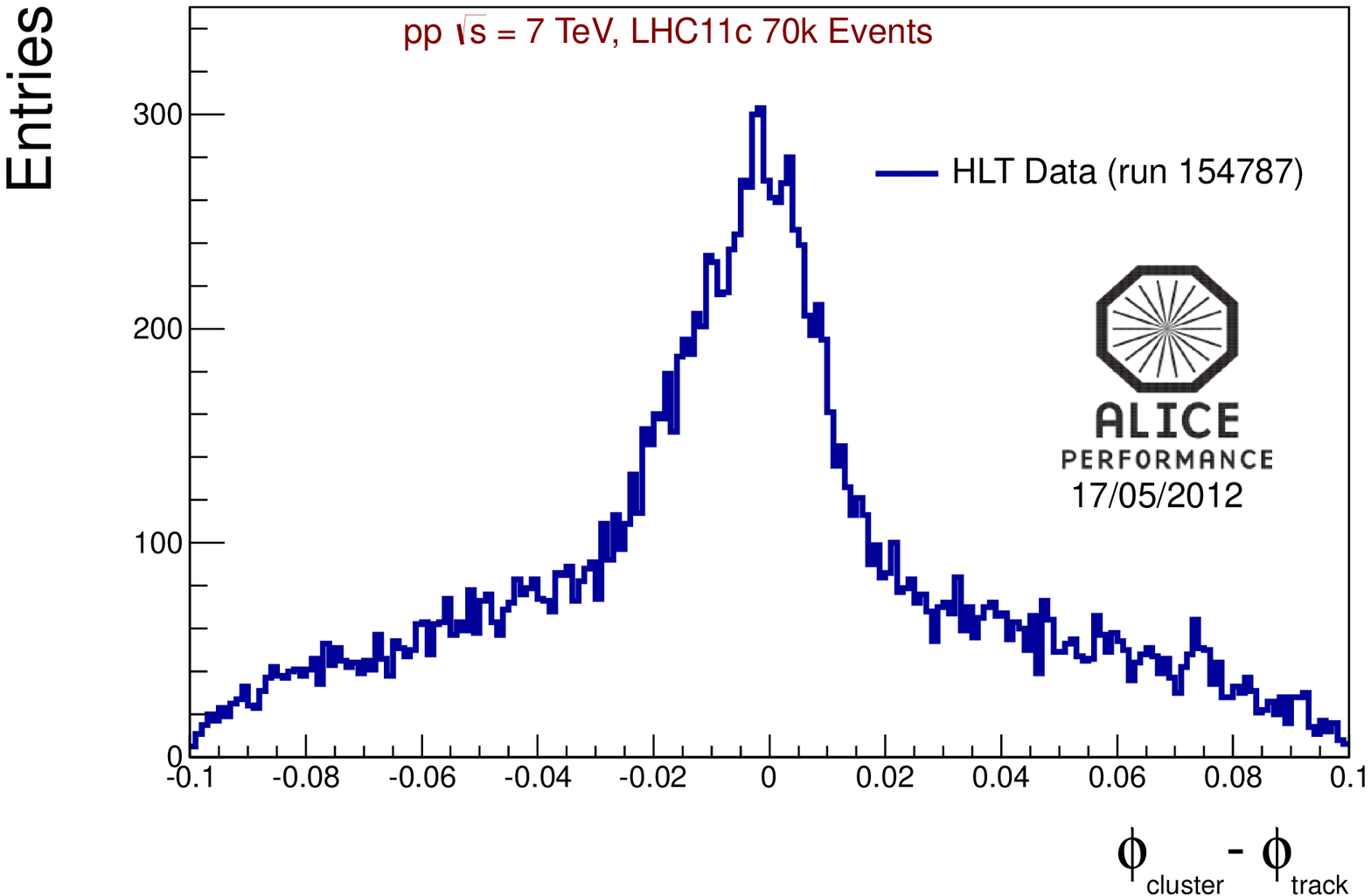}
\caption{\label{f7} 
Distribution of the residuals in azimuth ($\Delta\phi$) for the EMCal cluster and central barrel tracks
obtained using the HLT online chain for run 154787 (LHC11c), ~ 70 k events reconstructed. 
}
\end{minipage}\hspace{2pc}
\begin{minipage}{18pc}
\includegraphics[width=18pc]{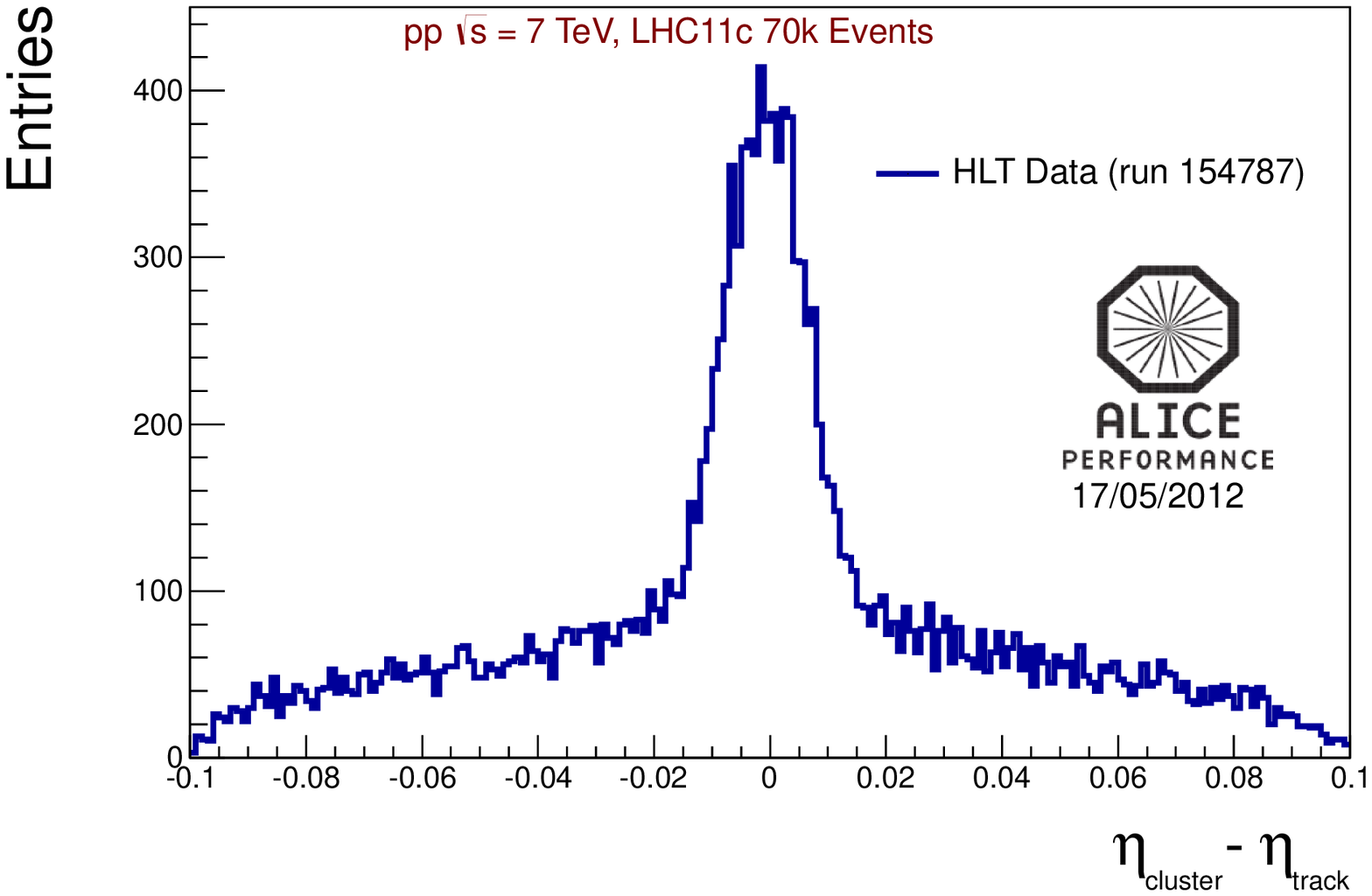}
\caption{\label{f8} 
Distribution of the residuals in pseudo-rapidity ($\Delta\eta$) for the EMCal cluster and central barrel tracks
obtained using the HLT online chain for run 154787 (LHC11c), ~ 70 k events reconstructed.
}
\end{minipage} 
\end{figure}

For this trigger  the cluster information reconstructed online by the EMCal HLT analysis 
chain is combined with the central barrel tracking information to produce complex event selection 
as a single electron trigger (matching of one extrapolated track with an EMCal cluster.
Performance and accuracy studies of the track matching component developed for this purpose 
have been done using simulated and real data taken during the 2011 LHC running period. 
Results are shown in Figures \ref{f7} and \ref{f8} where the cluster - track residuals
in azimuth and pseudo-rapidity units are to be compared with a calorimeter cell size of 
$0.014\times 0.014$.

In addition to the extrapolation of the track from the central barrel
to the EMCal interaction plane and the matching with a compatible nearby cluster, 
the electron trigger component must finally perform particle identification 
to issue a trigger decision. The selection of electron candidates is done 
using the $E/pc$ information where the energy is measured from the
EMCal cluster and the momentum from the central barrel track.
The trigger component is initialized with default values
for the cut of $0.8< E/pc <1.3$. The default cuts are stored in the HLT
conditions database and can be overridden via command line arguments
at configuration time (usually at start of run).

\begin{figure}[ht]
\includegraphics[width=24pc]{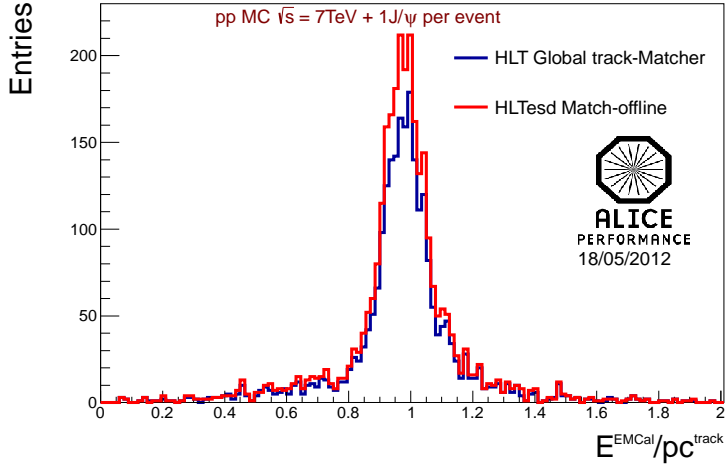}
\begin{minipage}[b]{12pc}\caption{\label{f9} 
$E/pc$ distributions obtained with the track extrapolation - cluster matching
via the online algorithms compared to the ESD-based tracking (red).}
\end{minipage}
\end{figure}

The performance of the electron trigger was studied using {\it pp} minimum 
bias data at 7 TeV with embedded $J/\Psi$ events.
Figure \ref{f9} shows the good agreement of the $E/pc$ distributions 
obtained with the track extrapolation - cluster matching 
performed using the online algorithms compared to the ESD-based tracking (red).

\begin{figure}[ht]
\includegraphics[width=24pc]{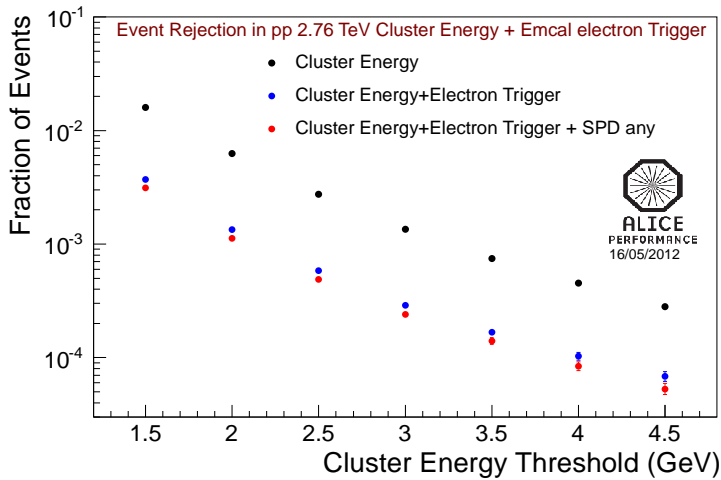}
\begin{minipage}[b]{12pc}\caption{\label{f10} 
Improvement in the event selection for $E_{e^-}>$~1~GeV from AliRoot simulation (anchor to LHC11b10a) with minimum bias {\it pp} at $\sqrt{s}=2.76$~TeV (EMCal full geometry).
The red points are obtained with the requirement of one hit in one of the silicon pixel (SPD) layers to reject a higher fraction of photon conversions.
}
\end{minipage}
\end{figure}

To determine the possible improvement of the event selection 
for electrons with energies above 1~GeV, AliRoot simulations of the HLT chain using LHC11b10a {\it pp} minimum bias data 
at 2.76 GeV and the EMCal full geometry (10 super-modules)  have been used. These studies
have shown that at least a factor 5 to 10 in event selection can be gained compared to the single shower trigger, as shown in Figure \ref{f10}.

\subsection{Jet trigger}

The EMCal online jet trigger component was developed to provide 
an unbiased jet sample by refining the hardware L1 trigger decisions.
In fact, the HLT post-processing can produce a sharper turn on curve 
using the track matching capabilities of the online reconstruction chain. 
In addition, a more accurate definition of the jet area than the one provided by the hardware L1 jet patch, 
can be obtained choosing a jet cone based on the jet direction calculated online.
The combination of the hadronic and electromagnetic energy provides a measurement of 
the total energy of the jet by matching the tracks identified as part of the jet with 
the corresponding EMCal neutral energy.

The use of the HLT jet trigger also allows a better characterization 
of the trigger response as a function of the centrality dependent threshold 
by re-processing the information from the V0 detector directly in HLT.

Performance considerations, due to the high particle multiplicity    
in {\it PbPb} collisions, impose that the track extrapolation is done only geometrically
without taking into account  multiple scattering effects
introduced by the material budget in front of the EMCal. 
The pure geometrical extrapolation accounts for a speedup factor of 20 in the
execution of the track matcher component with respect to the 
full-fledged track extrapolation used in {\it pp} collisions.

The identification of the jet tracks is performed using the anti-$k_T$ 
jet finder provided by the FastJet package.

The EMCal jet trigger was only partially tested during the 2011 data taking period
and will be fully commissioned for the LHC {\it pPb} run period in 2012.

\section{Monitoring components}

The role of the EMCal HLT reconstruction in {\it pp} collisions is targeted mainly on 
the monitoring functions since the expected event sizes are small enough for 
the complete collision event to be fully transferred to permanent storage. 
\begin{figure}[h]
\includegraphics[width=26pc]{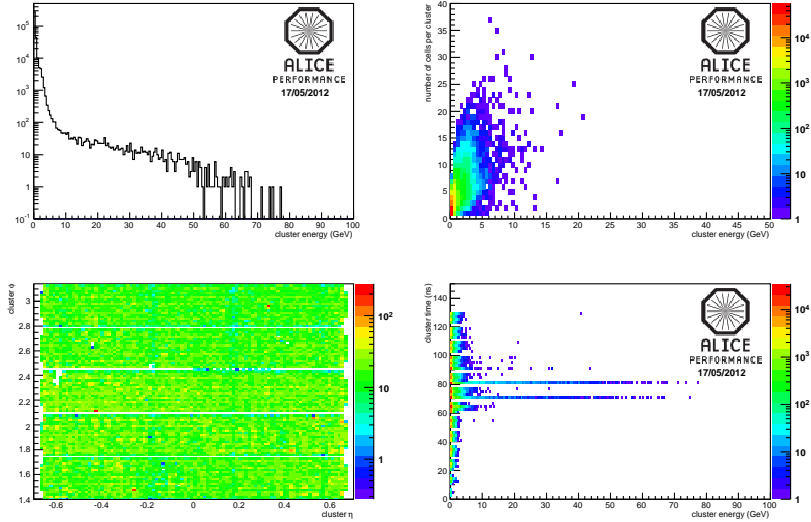}
\begin{minipage}[b]{12pc}
\caption{\label{f2} Output from the EMCal HLT monitoring component. Top left: cluster energy spectra as a function of the  reconstructed cluster energy; 
bottom left: cluster position in  $\eta$ and $\phi$ coordinates; bottom right: cluster time distribution; top right: number of cells per cluster vs cluster 
energy. LHC11b period, $\sqrt{s}=7$ TeV {\it pp} data, 10~kEvent analyzed.}
\end{minipage}
\end{figure}

In this respect, two monitoring components have been developed and deployed in the online chain.
The first component currently monitors reconstructed quantities, such as the cluster energy spectra and timing, 
the cluster position in the $\eta$ and $\phi$ coordinates, and the number of cells per cluster 
as a function of the cluster reconstructed energy as shown in Figure \ref{f2}.

\begin{figure}[h]
\includegraphics[width=24pc]{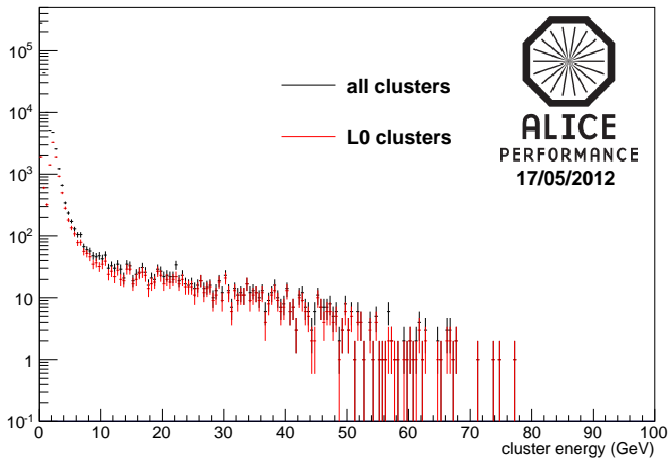}
\begin{minipage}[b]{14pc}
\caption{\label{f3} Energy spectrum for all clusters reconstructed by the EMCal (black points) superposed with the triggered cluster spectrum
(i.e. clusters reconstructed which also carry the L0 hardware trigger bit set, red points). }
\end{minipage} 
\end{figure}

The second component re-evaluates the EMCal hardware trigger decisions by recalculating
the cluster energy spectrum for all the clusters with the L0 trigger bit set as shown 
in Figure \ref{f3}. The L0 turn on curve can then be calculated online as the ratio between
the triggered and the reconstructed cluster spectra and monitored for the specific run.

No recalculation of hardware L1 trigger primitives was possible
during the 2011 data taking since the optical link from the EMCal L1 trigger unit 
could only installed during the 2011-2012 winter shutdown of the LHC hence
the software development for the L1 trigger monitoring is still underway.

\section{Conclusions}
In {\it pp} collisions the bandwidth to mass storage is sufficiently large to allow the recording of the full data volume at the maximum  
event rate at which ALICE can be read out, so rejection of the accepted events at the HLT level is not needed.
In this scenario the primary use of the EMCal HLT online chain relies on its monitoring
capabilities as discussed in the present report.

On the other hand, in {\it PbPb} collisions, online post-filtering in the HLT is foreseen 
to provide an additional event rejection and to sharpen the jet trigger response at the threshold.
A sharper trigger turn-on would be also relevant in {\it pp} running since it would lead 
to an optimization of the computing resources needed for the offline reconstruction.
Finally, the EMCal HLT jet trigger is expected to reduce the bias with respect to the hardware jet patch algorithm. 
Studies are still underway to quantify the response of the EMCal HLT trigger chain for the different collision modes.

\end{document}